\documentclass[aps, twocolumn, pra, showpacs, superscriptaddress]{revtex4}
\usepackage{amsfonts}
\usepackage{mathrsfs}
\usepackage{graphicx}
\usepackage{dcolumn}
\usepackage{bm}

\begin{document}
\title{Quantum entanglement of particles on a ring with fractional statistics}
\author{Hongli Guo}%

\affiliation{Institute of
Physics, Chinese Academy of Sciences, Beijing 100190, China}
\author{Yajiang Hao}
\affiliation{Department of Physics, University of Science and
Technology Beijing, Beijing 100083, China}
\author{Shu Chen}
\email{schen@aphy.iphy.ac.cn} \affiliation{Institute of Physics,
Chinese Academy of Sciences, Beijing 100190, China}
\date{\today}

\begin{abstract}
In this paper we investigate the von Neumann entropy in the ground
state of one-dimensional anyonic systems with the repulsive
interaction. Based on the Bethe-ansatz method, the entanglement
properties for the arbitrary statistical parameter
($0\leq\kappa\leq1$) are obtained from the one-particle reduced
density matrix in the full interacting regime. It is shown that the
entanglement entropy increases with the increase in the interaction
strength and statistical parameter. The statistic parameter affects
the entanglement properties from two aspects: renormalizing of the
effective interaction strength and introducing an additional anyonic
phase. We also evaluate the entanglement entropy of hard-core anyons
for different statistical parameters in order to clarify solely the
effect induced by the anyonic phase.
\end{abstract}

\maketitle


{\label{sec:level1}}
\section{Introduction}
In recent years quantum entanglement has attracted more and more
attentions because it is not only a key resource in quantum
information theory but also an essential concept in condensed matter
physics \cite{book}. It is well known that a great deal of
properties of quantum many-body systems are closely related with the
entanglement between particles, and the study of which is very
important for both the system composed of distinguishable particles
and the system composed of identical particles. However, unlike the
system of distinguishable particles, for which there are various
quantities to define and measure entanglement, the definition and
quantification of entanglement between identical particles is still
not very clear. Fortunately, the von Neumann entropy introduced by
reduced density matrix is a good quantity to describe quantum
entanglement for systems consisting of identical particles
\cite{Law, You, Zhou, pra66042113,prb63085311,pra64022303,Long}. The
entanglement entropy of two identical bosons or fermions in abstract
wave functions has been studied by the Schmidt decomposition
\cite{You,Long,prb63085311,pra64022303}, while the entanglement
between two identical interacting trapped atoms in a continuous
system was studied in Refs.\cite{Law, Zhou, pra66042113}. It is
shown that the statistical properties of identical particles play
very important roles in their entanglement behaviors.

As a natural generalization of boson and fermion, anyon was proposed
to describe the particle obeying fractional statistics and has been
a subject of great interest in the past decades
\cite{book2,laughlin1,laughlin2, prl661529,prl521583,2D1,2D2}.
Although the concept of anyon arises originally from two-dimensional
systems \cite{prl521583,2D1,2D2} related to fractional quantum Hall
effect (FQH) and high-temperature superconductivity
\cite{book2,laughlin1,laughlin2}, the study of 1D anyonic model has
attracted great theoretical interest
\cite{prl661529,Kundu,1D1,1D2,Wang,Girardeau,Patu,del
Campo,junction,Santachiara08,Guan2,Guan3,Guan1,Hao1,Hao2,Calabrese09}.
Motivated by possible experiments with cold atoms to simulate the
creation and manipulation of anyons
\cite{Paredes_anyon,Aguado,Jiang}, and the possibility of performing
topological quantum computation \cite{tqc}, the properties of 1D
anyons are under current research focus. Several studies have been
devoted to 1D anyons with a $\delta$-function potential interaction
\cite{Kundu,Guan1,Guan2,Guan3,Hao1,1D1}  and the limiting cases of
hard-core anyons \cite{1D2,Santachiara08,Girardeau,Patu,del
Campo,Calabrese09,Hao2}. Particularly, the 1D interacting anyon
model is exactly solvable by the Bethe-ansatz method as firstly
found by Kundu \cite{Kundu}. Despite the intensive studies, the
entanglement properties in anyonic systems are rarely studied except
for the model in the hard-core limit \cite{1D2}.

In this work, we investigate the entanglement properties of a
continuous system composed of $N$ anyonic particles with repulsive
contact interaction on a ring of length $L$ by calculating the von
Neumann entropy of the single-particle reduced density matrix. In
general, it is hard to calculate the von Neumann entropy of a
continuous many-body system analytically. So far, most of the
studies focus on the two-particle system \cite{Law, Zhou,
pra66042113}. The integrability of the exactly solvable many-body
system provides us the possibility to study the entanglement
properties of a many-body system analytically. Based on the Bethe
ansatz solution of the interacting anyonic model \cite{Kundu,Guan1},
we evaluate the one-particle reduced density matrix firstly and then
obtain the von Neumann entropy for the arbitrary statistical
parameter in the full interacting regime.

This paper is organized as follows. In section II, we first give a
brief introduction to the model and formulate the method. In section
III, we first consider the Bose limit and focus on the effect of
interacting strength on the von Neumann entropy. In section IV, we
deal with the general anyonic case and discuss the effect of
statistical parameter $\kappa$ on the entanglement properties by
calculating the von Neumann entropy for different $\kappa$. A brief
summary is given in Section V.

\section{models and methods}
The second quantized Hamiltonian for the one-dimensional anyonic
system is formulated as
\begin{eqnarray}
\mathcal{H}_A &=&-\frac{\hbar ^2}{2m}\int_0^Ldx \Psi _A^{\dagger }
\frac{\partial^2 }{\partial x^2}\Psi _A \nonumber \\
&&+\frac {g_{1D}}{2} \int_0^Ldx\Psi _A^{\dagger }\left( x\right)
\Psi _A^{\dagger }\left( x\right) \Psi _A\left( x\right) \Psi
_A\left( x\right), \label{H}
\end{eqnarray}
in which $m$ is the mass of anyons and $g_{1D}$ denotes interacting
strength between anyons \cite{Kundu,Guan1}. Here the field operators
obey anyonic commutation relations $ \Psi ^{\dag }_{A}(x_{1})\Psi
^{\dag }_{A}(x_{2})= e^{i\kappa\pi\epsilon (x_{1}-x_{2})}\Psi ^{\dag
}_{A}(x_{2})\Psi ^{\dag}_{A}(x_{1})$, and $ \Psi _{A}(x_{1})\Psi
^{\dag }_{A}(x_{2})=\delta (x_{1}-x_{2})+e^{-i\kappa \pi \epsilon
(x_{1}-x_{2})}\Psi ^{\dag }_{A}(x_{2})\Psi _{A}(x_{1})$ with
$\epsilon(x-y)=1, -1, 0$ for $x>y, x<y$, and $x=y$ respectively. The
model (\ref{H}) is known to be exactly solvable \cite{Kundu}. The
parameter $\kappa$ characterizes the statistical property of the
anyonic system with $\kappa =0$ and $\kappa =1.0$ corresponding to
Bose statistics and Fermi statistics respectively. The dependence of
entanglement properties on both the interaction constant $g_{1D}$
and statistical parameter $\kappa$ ($0\leq\kappa\leq1$) will be
considered.

The eigenvalue problem of Hamiltonian (\ref{H}) can be reduced to
the quantum mechanical problem of $N$ anyons with $\delta$
interaction \cite{Kundu,Guan1}
\begin{eqnarray}
H\psi(x_{1},...,x_{N})=E\psi(x_{1},...,x_{N})
\end{eqnarray}
with
\begin{eqnarray}
H=-\sum\limits_{i=1}^{N}\frac{\partial ^{2}}{\partial x_{i}^{2}}
+2c\sum\limits_{1\leq i\leq j\leq N}^{N}\delta
(x_{i}-x_{j}),\label{H2}
\end{eqnarray}
where the natural unit is used and $c=mg_{1D}/\hbar^2$ ($c>0$) is a
dimensionless interaction constant.

In terms of the exact ground-state wavefunction $\psi$, the
single-particle reduced density matrix is defined as
\begin{equation}
\hat{\rho}_1 = \mathrm{Tr}_{2,3,...,N}|\psi\rangle \langle \psi|,
\end{equation}
where the trace means to do integrations over all the position
coordinates except one of them. The single-particle entanglement is
quantified by the von Neumann entropy as
\begin{equation}
S =-\mathrm{Tr} (\hat{\rho}_1 \log _{2} \hat{\rho}_1) ,
\end{equation}%
where $\hat{\rho}$ is the one-particle reduced density matrix with
the normalization condition Tr$\hat{\rho} =1$.
In coordinate representation, the one-particle reduced density
matrix is expressed as
\begin{eqnarray}
& & \rho_1(x,x^{\prime}) \nonumber \\
&=& \frac{\int_{0}^{L}dx_{2}...dx_{N}
[\psi^{\ast}(x,x_{2},...x_{N})\psi(x^{\prime},x_{2},...,x_{N})]}
{\int_{0}^{L}dx_{1}...dx_{N}|\psi(x_{1},x_{2},...,x_{N})|^{2}}.~~~~
\label{density matrix}
\end{eqnarray}
Obviously the (\ref{density matrix}) is Hermitian, i.e., we have
$\rho_1^{\ast}(x,x^{\prime})=\rho_1(x^{\prime},x)$. The eigen-
equation of the one-particle reduced density matrix $(\ref{density
matrix})$ is
\begin{eqnarray}
\int_{0}^{L}dx^{\prime}\rho_1(x,x^{\prime})\phi_{\eta}(x^{\prime})=
\lambda_{\eta}\phi_{\eta}(x),\label{eigenquation}
\end{eqnarray}
where $\lambda_{\eta}$ are the occupation numbers for natural
orbitals $\phi_{\eta}(x)$ which form a complete and orthonormal set
of functions. The one-particle reduced density matrix is diagonal in
the basis of natural orbitals and
$\sum_{\eta=1}^{\infty}\lambda_{\eta}=1$. In terms of eigenvalues
$\lambda_{\eta}$ and the eigenfunctions $\phi_{\eta}(x)$, we can
rewrite
\[
\rho_1(x,x^{\prime})=\sum_{\eta=1}^{\infty}\lambda_{\eta}
\phi_{\eta}(x) \phi_{\eta}^{\ast}(x^{\prime})
\]
and the von Neumann entropy then reads
\begin{equation}
S=-\sum_{\eta=1}^{\infty}\lambda_{\eta}\log_{2}\lambda_{\eta} .
\end{equation}

From the above scheme, we can understand the difficulties which
prevent us from studying the entanglement properties of a many-body
system. First, the calculation of the ground-state wavefunction for
a many-body system is generally difficult except for some exactly
solvable systems. Furthermore, even though the exact many-body wave
function is constructed, the calculation of the reduced density
matrix for a large system remains a difficult task due to the time
consuming to calculate multidimensional integrals.

{\label{sec:level1}}
\section{The dependence of entanglement on the interaction strength}
In this section, we shall focus on the Bose limit and study the
dependence of entanglement on the interaction strength. The effect
of fractional statistics will be discussed in the next section.  In
the Bose limit ($\kappa=0$), the model is reduced to the well-known
Lieb-Linger model \cite{1963}. In this case the field operators
$\Psi_A ^{\dag }(x)$ and $\Psi_A (x)$ obey boson commutation
relations and the wavefunction $\psi(x_{1},...,x_{N})$ satisfies
exchange symmetry. In the scheme of Bethe-ansatz method \cite{1963},
the many-particle wave function can be formulated as
\begin{eqnarray}
\psi(x_{1},...,x_{N}) &=&\sum_{Q}\theta(x_{q_{N}}-x_{q_{N-1}})
...\theta(x_{q_{2}}-x_{q_{1}})\nonumber\\
&&\times
\varphi_{Q}(x_{q_{1}},x_{q_{2}},...,x_{q_{N}}),\label{wavefunction}
\end{eqnarray}
where $Q$ labels the region $0\leq x_{q_{1}}\leq x_{q_{2}}...\leq
x_{q_{N}}\leq L$, in which $q_{1},q_{2},...,q_{N}$ is one of the
permutations of $1,2,...,N$, $\Sigma_{Q}$ sums over all permutations
and $\theta(x-y)$ is the step function. Here
$\varphi_{Q}(x_{q_{1}},x_{q_{2}},...,x_{q_{N}})$ takes the
Bethe-ansatz type
\begin{eqnarray}
\varphi_{Q}(x_{q_{1}},...,x_{q_{N}})=\sum\limits_{P}[A_{p_{1}p_{2}...p_{N}}\exp(i\Sigma_{j}(k_{p_{j}}x_{q_{j}}))]\label{wavefunction1}
\end{eqnarray}
with
$A_{p_{1}p_{2}...p_{N}}=\varepsilon_{P}\prod_{j<l}^{N}(ik_{p_{l}}-ik_{p_{j}}+c)$
and ${k_{p_{j}}}$ is a set of quasi-momentums determined by the
Bethe-ansatz equations. Here $p_{1},p_{2},...,p_{N}$ means one of
permutations of $1,2,...,N$, and $\varepsilon_{P}$ denotes a $+$
$(-)$ sign associated with even (odd) permutation of $P$. Using the
periodical boundary condition, we can obtain Bethe-ansatz equations
\cite{1963} whose logarithmic forms are formulated as
\begin{eqnarray}
k_{j}L=2n_{j}\pi-\sum\limits_{l=1(l\neq j)}^{N}2\arctan
\Big(\frac{k_{j}-k_{l}}{c}\Big),\label{bae}
\end{eqnarray}
where ${n_{j}}$ is a set of integers to determine the eigenstates
and for the ground state $n_{j}=(N+1)/2-j$ $(1\leq j\leq N)$. The
energy of the system is $E=\sum_{j=1}^{N}k_{j}^{2}$ and the total
momentum is $k=\sum_{j=1}^{N}k_{j}$.
\begin{figure}
\includegraphics[height=6cm,width=\linewidth]{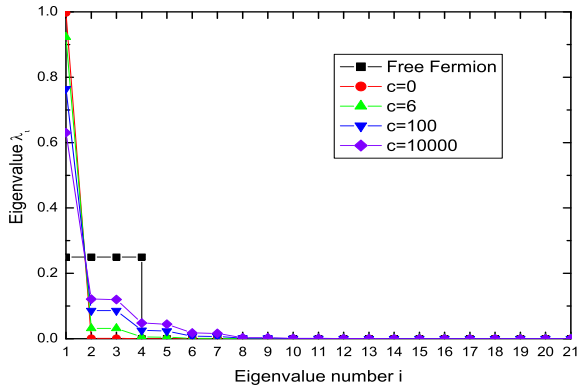}
\caption{\label{fig:epsart} The occupation numbers of the
interacting Bose system for different interaction strength $c$.}
\end{figure}
\begin{figure}
\includegraphics[height=7cm,width=\linewidth]{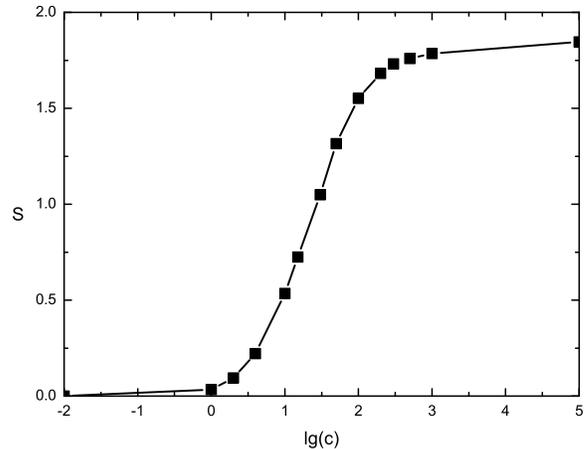}
\caption{\label{fig:epsart} The entanglement entropy $S$ versus the
logarithm of the repulsive interaction constant $c$ for a system
with $4$ particles. One can see that $S$ monotonically and smoothly
increases from zero to about 1.846 with the growth of $c$ from zero
to the infinity limit.}
\end{figure}

By numerically solving the Bethe-ansatz equations (\ref{bae}), we
can obtain the exact ground-state wavefunction, and then the
one-particle reduced density matrix (\ref{density matrix}). Solving
the eigenvalue problem (\ref{eigenquation}) numerically, we can get
a series of $\lambda_i$ and then determine the entanglement entropy.
For simplicity, we shall discuss the many-body system with $N=4$ in
the following context. In Fig. 1, we show the occupation numbers for
the interacting Bose system composed of four identical bosons versus
different interaction strength $c$. At $c=0$, only the lowest nature
orbital is occupied which means all the bosons condensate to the
ground state. The occupation number of the lowest natural orbital
$\lambda_1$ decreases with the increase of the interaction strength,
accompanying with the increase of occupation numbers of higher
natural orbitals. Our numerical results for the dependence of
ground-state entanglement $S$ on the interaction strength are shown
in Fig. 2.
In order to exhibit the change with $c$ in a wide range, the
logarithm coordinate for c is used in this figure. It is shown that
the entanglement entropy changes monotonically with the change of
interaction constant.
When there is no interaction between bosons ($c=0$), no entanglement
exists in the system. Along with the growth of interaction constant,
the entanglement entropy $S$ increases slowly in the weakly
interacting regime, and then goes sharply to $1.74$. When the
interaction gets close to the strongly interacting regime, the
entanglement entropy $S$ slowly approaches to about $1.846$, which
is smaller than $2$. Our result is accordant with \cite{You,Zhou},
in which we note the entanglement entropy of $N$ identical
boson-particle system ranges from  $S=0$ for free-boson state to a
maximum $S$ in the infinitely repulsive limit which is smaller than
$S=\log_2N$. This can be understood as follows: when there is no
interaction between particles, no correlation exists between bosons
and the occupation number of lowest-energy state is $N$; while in
the strong interaction limit ($c\rightarrow\infty$), particles will
be prevented from occupying the same state and higher-energy states
should be occupied.

{\label{sec:level1}}
\section{The dependence of entanglement on the anyonic parameter}
Now we turn to the dependence of ground state entanglement entropy
on the statistics. For anyonic system the many-body wave function
shall satisfy the generalized symmetry
\begin{eqnarray}
\psi(...,x_{i},...,x_{j},...) =
e^{-i\theta}\psi(...,x_{j},...,x_{i},...),
\end{eqnarray}
where the anyonic phase
\[
\theta=\kappa
\pi\left[\sum_{k=i+1}^{j}\epsilon(x_{i}-x_{k})-\sum_{k=i+1}^{j-1}\epsilon(x_{j}-x_{k})\right].
\]
for $i<j$. Considering the symmetry under coordinates reflection, we
confine $\kappa$ to $[0,1]$ in the present paper. The wavefunction
of anyons takes a similar form with that of bosons ($\kappa=0$)
\cite{Kundu,Guan1}
\begin{eqnarray}
\psi_{A}(x_{1},...,x_{N})&=&\sum_{Q}\theta(x_{q_{N}}-x_{q_{N-1}})
...\theta(x_{q_{2}}-x_{q_{1}})\nonumber\\
&&\times\phi_{A}\varphi_{Q}(x_{q_{1}},x_{q_{2}},...,x_{q_{N}}),\label{anyonicwf}
\end{eqnarray}
where $\phi_{A}$ is an additional anyonic phase part
\begin{eqnarray}
\phi_{A}=\exp(-i\frac{\kappa
\pi}{2}\sum_{q_i<q_j}\epsilon(x_{q_{i}}-x_{q_{j}}))
\end{eqnarray}
and $\varphi_{Q}(x_{q_{1}},x_{q_{2}},...,x_{q_{N}})$ has the same
form as that of Lieb-Liniger Bose gas (\ref{wavefunction1}) except
that now we have
$A_{p_{1}p_{2}...p_{N}}=\varepsilon_{P}\prod_{j<l}^{N}(ik_{p_{l}}-ik_{p_{j}}+c^{\prime
})$ with
\begin{eqnarray}
c^{\prime }=c/\cos(\kappa/2)\label{effective constant}.
\end{eqnarray}
Similarly, the quasi-momenta $k_i$ is determined by the Bethe ansatz
equations
\begin{eqnarray}
k_{j}L=2n_{j}\pi-\sum\limits_{l=1(l\neq j)}^{N}2\arctan
\Big(\frac{k_{j}-k_{l}}{c^{\prime}}\Big),\label{bae1}
\end{eqnarray}
under the twisted boundary condition $
\psi_{A}(0,x_{2},...,x_{N})=e^{i\kappa
\pi(N-1)}\psi_{A}(L,x_{2},...,x_{N}) $. The Bethe-ansatz equations
have the same form as that for Lieb-Liniger Boson gas (\ref{bae}) if
we replace $c$ with the renormalized interaction constant $c^{\prime
}$. According to (\ref{effective constant}) the effective
interaction between anyons depends on the statistical parameter
$\kappa$, which increases with the increase of $\kappa$ and
approaches $\infty$ in the fermionic limit ($\kappa \rightarrow
1.0$).

\begin{figure}
\includegraphics[height=6cm,width=\linewidth]{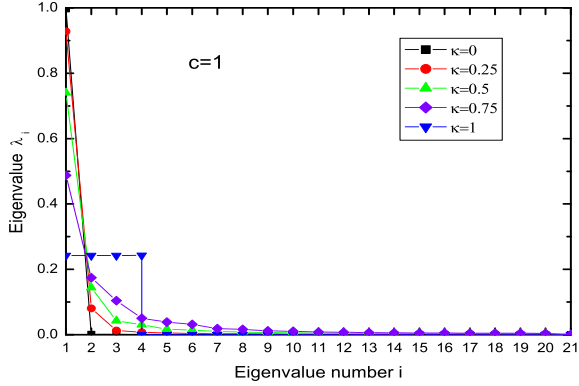}
\caption{\label{fig:epsart}The occupation numbers for different
anyonic parameter $\kappa$ with $c=1$.}
\end{figure}
\begin{figure}
\includegraphics[height=8cm,width=\linewidth]{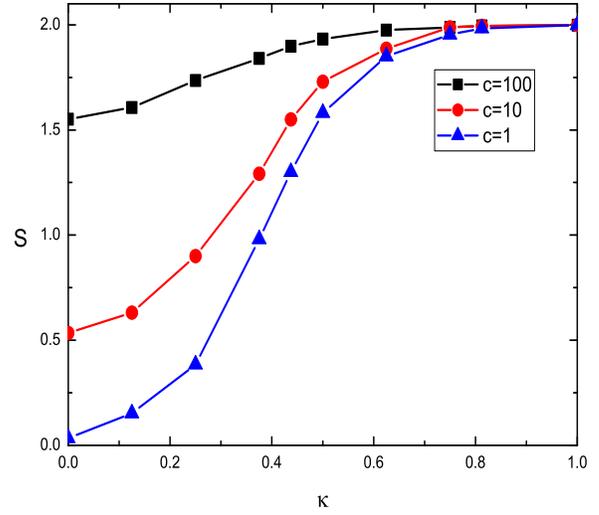}
\caption{\label{fig:epsart}(Color Online) The entanglement entropy
versus the statistics parameter $\kappa$ for the anyonic system with
$c=1$, $c=10$ and $c=100$, respectively.}
\end{figure}

With the same procedure as presented in the above section we obtain
the one-particle entanglement entropy for various c. As a concrete
example, in Figure 3 we display the change of occupation numbers
with different statistical parameters at a fixed interaction
strength $c=1$. The dependence of entanglement entropy on the
statistical parameters $\kappa$ is displayed in Figure 4 for $c=1$,
10 and 100. It is shown that the entanglement entropy increases with
the increasing of anyonic parameter $\kappa$ and interaction
strength $c$. For the case of $c=10$, as $\kappa=0$ the system
reduces to Lieb-Liniger gas and the entanglement entropy $S$ is
about $0.535$; while as $\kappa=1/2$, $S$ increases to about $1.730$
and reaches $2$ in the Fermi limit with $\kappa=1.0$. The similar
behaviors are displayed for $c=1$ and $c=100$. For different $c$ the
entanglement entropy $S$ converges to $\log_2N$ when $\kappa=\pi$
($S_{max}=2.0$ here).

In order to understand why and how $\kappa$ affects the entanglement
entropy, we compare the wavefunction of anyonic gas with that of
Bose gas. According to (\ref{effective constant}) and (\ref{bae}),
it is easy to find that the set of quasi-momenta {$k_{j}$} depend on
$c^{\prime }$ and the wave function $\varphi_{Q}$ changes along with
the change of $\kappa$. In addition the phase factor $\phi_{A}$
includes the statistical parameter obviously. So the influence of
statistical parameter $\kappa$ on entanglement contains two parts:
one coming from the phase $\phi_{A}$ and the other coming from
wavefunction $\varphi_{Q}$ through the renormalization of effective
interaction strength $c^{\prime }$.

\begin{figure}
\includegraphics[height=7cm,width=\linewidth]{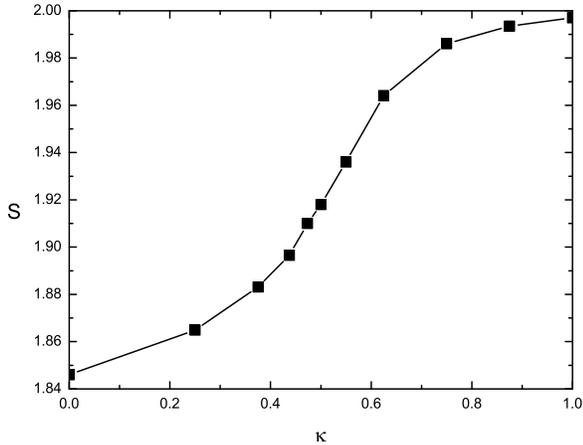}
\caption{\label{fig:epsart} The entanglement entropy versus the
statistics parameter $\kappa$ for the hard-core anyonic gas  with
$N=4$.}
\end{figure}

Finally in order to clarify solely the dependence of entanglement
entropy on the part of anyonic phase $\phi_A$, we investigate the
hard-core anyons, which can be studied using the anyon-fermion
mapping method \cite{1D2,Girardeau}. In this situation the effective
interaction $c^{\prime }=\infty$ and the system has the same set of
quasi-momenta ${k_{j}}$ whatever the statistical parameter $\kappa$
is. The fractional statistics only contributes a phase $\phi_{A}$ in
the wavefunction. According to the numerical result shown in Figure
5, the entanglement entropy of hard-core anyonic gas increases
monotonically when the statistical parameter $\kappa$ changes from
the Bose limit $\kappa=0$ to the Fermi limit $\kappa=1.0$.

{\label{sec:level1}}
\section{Summary}
In summary, we have investigated the ground-state entanglement of
the 1D anyonic system with repulsive interaction by calculating the
one-particle von Neumann entropy in the full interacting regime
($0\leq c\leq \infty$) for arbitrary statistical parameter ($0\leq
\kappa \leq 1.0$). Using the Bethe-ansatz method, we obtain the
exact ground-state wavefunction, and thus the one-particle reduced
density matrix and the von Neumann entropy. In the Bose limit
($\kappa=0$) the entanglement entropy increases monotonically with
the increase of the interaction strength and approaches a maximum in
the hard-core limit. While for anyonic system it is shown that the
entanglement entropy increases monotonically both with the increase
in the interaction strength and statistical parameter. The anyonic
gas gets to the maximum entanglement entropy $\log_2N$ in the Fermi
limit ($\kappa=1.0$). The statistical parameter $\kappa$ affects the
entanglement properties of the anyonic system by renormalizing the
effective interaction strength and introducing an additional anyonic
phase in the wavefunction. The influence of anyonic phase on the
entanglement is also clarified by evaluating the entanglement
entropy of hard-core anyons for different statistical parameters.

{\label{sec:level1}}
\begin{acknowledgments}
The authors would like to thank Z. Liu and D. L. Zhou for helpful
discussions. This work is supported by NSF of China under Grant No.
10821403 and No. 10847105, programs of Chinese Academy of Sciences,
National Program for Basic Research of MOST.
\end{acknowledgments}

\bibliography{apssamp}
\end{document}